%% file: main.tex
\documentclass[acmsmall,screen]{acmart}

\usepackage{booktabs}
\usepackage{multirow}
\usepackage{graphicx}
\usepackage{listings}
\usepackage{xspace}
\usepackage{pifont}
\usepackage{microtype}
\usepackage[most]{tcolorbox}
\usepackage{xcolor}
\usepackage{mdframed}

\AtBeginDocument{%
  }

\renewcommand\footnotetextcopyrightpermission[1]{}

\begin{document}


\title{Making Small Code LLMs More Efficient for Secure Code Review with Neuro-Symbolic Fine-Tuning}

\author{Mostafijur Rahman Akhond}
\affiliation{%
  \institution{York University}
  \city{Toronto}
  \country{Canada}}
\email{mostafij@yorku.ca}

\author{Mahmuda Khatun}
\affiliation{%
  \institution{York University}
  \city{Toronto}
  \country{Canada}}
\email{mahmuda@yorku.ca}

\author{Gias Uddin}
\affiliation{%
  \institution{York University}
  \city{Toronto}
  \country{Canada}}
\email{guddin@yorku.ca}



\renewcommand{\shortauthors}{Akhond et al.}

\newcommand{\todo}{{\color{red}\bfseries TODO}}
\newcommand{\cmark}{\ding{51}}
\newcommand{\xmark}{\ding{55}}

\newcommand{\ph}[1]{\textcolor{black}{#1}}
\newcounter{observation}
\definecolor{obsmain}{HTML}{1F4E79}
\definecolor{obsbg}{HTML}{EEF3F9}

\newcommand{\observation}[1]{%
  \refstepcounter{observation}%
  \begin{tcolorbox}[
    enhanced, breakable=false,
    colback=obsbg, colframe=obsmain,
    boxrule=0.4pt, arc=2pt,
    left=6pt, right=6pt, top=7pt, bottom=3pt,
    before skip=10pt, after skip=6pt,
    borderline west={2.5pt}{0pt}{obsmain},
    attach boxed title to top left={yshift=-2.2mm, xshift=6pt},
    boxed title style={colback=obsmain, colframe=obsmain, arc=2pt,
      boxrule=0pt, left=4pt, right=4pt, top=1pt, bottom=1pt},
    title={\small\bfseries\textcolor{white}{Observation~\theobservation}}]
  #1
  \end{tcolorbox}%
}


\input{body/abstract}

\keywords{Code Security, LLM, Security Monologues, CWE, Security Semantics }

\received{** **** ****}
\received[revised]{** **** ****}
\received[accepted]{** **** ****}

\maketitle
\pagestyle{plain}

\input{body/introduction}

\input{body/prilimineries}
\input{body/data_collection}

\input{body/study_design}

\input{body/resullt}
\input{body/threads}
\input{body/discussion}
\input{body/related_works}

\input{body/conclusion}

\section*{Data Availability} 
A replication package is available at: \url{https://anonymous.4open.science/r/secSemScan-2615.}

\bibliographystyle{ACM-Reference-Format}
\bibliography{main}

\appendix



\end{document}

%% file: body/abstract.tex
\begin{abstract}

As AI coding assistants generate more code, the need to review it for security flaws grows. Existing approaches leave a gap: static analyzers identify program sites but offer little explanation, while language-model explanations can be difficult to verify. Small open-weight code models could make security review available without sending proprietary code to a hosted service. We study whether fine-tuning helps these models turn security evidence into explanations that developers and tools can check. We present secSemScan, which pairs a static analyzer with a locally deployable model that writes a schema-constrained security monologue naming program sites, assigning a CWE, and giving a machine-parseable verdict. We also introduce OGSM (Oracle-Grounded Security Monologue), a dataset of 1,993 C/C\texttt{++} functions covering 22 CWEs in the taint and lifetime families. Each record pairs a function with analyzer-derived or fix-diff-derived evidence and a monologue. We fine-tune three open-weight code models in two stages, first on the output format and then on the monologue content. Because evaluation prompts supply the verdict and evidence, our results measure how well models use that information, not whether they detect vulnerabilities independently. On 355 test records, fine-tuning raised Qwen2.5-Coder-7B's verdict accuracy from 4.8\% to 96.6\%; DeepSeek-Coder-6.7B reached 78.9\%, and StarCoder2-7B reached 43.7\%. Qwen2.5-Coder-7B cited 87.5\% of Tier~1 oracle sites, and every line it cited belonged to the oracle evidence. The results suggest that fine-tuning helps small models produce more structured, evidence-linked security reasoning.

\end{abstract}

%% file: body/introduction.tex
\section{Introduction}
\label{sec:intro}

AI coding assistants can speed up development~\cite{ziegler2022productivity, peng2023impact}, but the code they generate can contain vulnerabilities~\cite{pearce2022asleep, he2023large}. Coding agents add to the review workload: a study of 128,018 GitHub projects estimates that 22.20\%--28.66\% had adopted them and reports more vulnerabilities in projects using agents than in projects developed without them~\cite{robbes2026agentic}. Reviewing generated code is becoming a bottleneck~\cite{zhang2026ccrab}. At the same time, many organizations are reluctant to send proprietary code to third-party services~\cite{klemmer2024using, liang2024large, rzanny2026decision}. These pressures motivate our question: can a relatively small, open-weight code model produce security explanations that developers can check and run locally?

We consider two requirements for such a reviewer. Its findings must be \emph{checkable} (D1): each finding identifies the program sites that support it, so a developer or a tool can inspect the claim. It must also be \emph{local} (D2): the model can run on hardware controlled by the organization, keeping proprietary code in-house. Existing approaches tend to meet one requirement more clearly than the other. Static analyzers run locally and report code locations, but cover only the weaknesses encoded in their rules and often provide little explanation for triage~\cite{lipp2022effectiveness}. Learned detectors can also run locally, but commonly return a function-level label or suspicious lines rather than a trace of the sites needed to check a finding~\cite{fu2022linevul, du2024vulllm}. Hosted LLMs can explain suspected vulnerabilities in prose~\cite{zhang2023prompt, khare2023understanding}, but their explanations may be unfaithful~\cite{ullah2024llms}, and using them requires sending code to an external service. Neuro-symbolic systems combine LLMs with program analysis~\cite{sun2024gptscan, li2025iris, li2024llift, wang2024llmdfa}; locally deployable models trained on teacher-generated reasoning take another route~\cite{weyssow2025r2vul, du2024vulllm}. These lines of work leave open whether small, locally deployable models can be trained to express security evidence in a form that a tool can check.

We study two weakness families, \emph{taint} and \emph{lifetime}, because each can be characterized by a small tuple of program sites that a static analyzer can verify independently of the model. Both are established security problems: taint-style vulnerabilities are a common target of program analysis~\cite{tsipenyuk2005seven, yamaguchi2014modeling}, while memory-safety vulnerabilities include temporal errors such as use-after-free~\cite{szekeres2013sok}. The two families also require different forms of program reasoning. For taint weaknesses such as CWE-78 and CWE-120, the evidence follows data flow and can be represented as a (source, sink) trace~\cite{yamaguchi2014modeling}; for lifetime weaknesses such as CWE-416, it follows object lifetime through an (alloc, free, use) trace~\cite{szekeres2013sok}. A static analyzer can produce and verify these traces, but a bare trace is difficult for a developer to triage, while free-form prose is difficult for a tool to check. We therefore need an output that preserves this verifiable program evidence while presenting it as an explanation useful to developers.

We present secSemScan, a neuro-symbolic approach in which the analyzer supplies the evidence and a locally deployable model turns it into a readable, checkable explanation. Semgrep~\cite{semgrep} with custom taint and lifetime rules serves as the analyzer. We fine-tune an open-weight code model with LoRA~\cite{hu2022lora} to write a \emph{security monologue}, adapting monologue reasoning~\cite{ding2024semcoder}: a narration, constrained by a fixed schema, that names each site by identifier and line, assigns a CWE, and ends with a machine-parseable verdict. The symbolic component enters in three places. At data construction, the analyzer's trace, or the fix diff where no rule fires, is given to a generator LLM, so each narration names sites that come from the analyzer or the fix and not from the generator's own guess. At training and evaluation time, the prompt supplies the trace to the model. At inference, a fresh analyzer run checks the sites that a monologue names. We apply the approach to three open-weight models: Qwen2.5-Coder-7B~\cite{hui2024qwen25coder}, DeepSeek-Coder-6.7B~\cite{guo2024deepseekcoder}, and StarCoder2-7B~\cite{lozhkov2024starcoder2}.

We study two questions. The first is monologue \emph{creation}: can an open-weight model learn to write schema-conforming monologues that state the right verdict and CWE? The second is monologue \emph{usefulness}: do the monologues contain what D1 requires, namely the oracle's sites, identifiers that occur in the code, and a CWE that agrees with the verdict? We define usefulness by these checkable properties and do not measure developer effort or triage time. We evaluate on OGSM (Oracle-Grounded Security Monologue), a dataset we construct from DiverseVul, PairVul, and MVD: 1,993 C/C\texttt{++} functions (1,010 vulnerable, 983 safe) covering 22 CWEs, split by CVE into 1,248 training, 390 validation, and 355 test records. Each record pairs a function with a verdict, a CWE, evidence, and a security monologue, where the evidence is either an analyzer trace or the lines changed by a fix commit. Every prompt in the evaluation states the record's verdict, CWE, and oracle trace, so the evaluation measures whether a model turns a supplied trace into a narration.

On the 355 test records, fine-tuning raised verdict accuracy of Qwen2.5-Coder-7B from 4.8\% to 96.6\%, and the result is unchanged across five sampled outputs per record. The fine-tuned model cites 87.5\% of the Tier~1 oracle sites (untuned: 60.0\%), and every line it cites is an oracle site. The gain depends on the base model: verdict accuracy is 78.9\% for DeepSeek-Coder-6.7B and 43.7\% for StarCoder2-7B. Our checks test the surface of a narration, not whether its explanation is correct. A case study on five Linux kernel functions, in which Semgrep fires no rule, shows the model labeling all five VULNERABLE with the correct CWE. 


This paper makes three contributions. First, OGSM, a C/C\texttt{++} dataset in which each record pairs a function with a verdict, a CWE, evidence tagged by tier (analyzer-confirmed or diff-derived), and a security monologue. Second, a two-stage fine-tuning recipe that separates learning the schema from learning its content. Third, an evaluation of creation and usefulness, with rule-based checks that need no judge model. The design also checks a monologue's sites with a fresh analyzer run at inference, but the evaluation does not test this step, and whether the approach finds vulnerabilities that the analyzer's rules miss remains open.



%% file: body/prilimineries.tex
\section{Preliminaries}
\label{sec:prelim}


\subsection{Motivating Example}
\label{sec:prelim-example}

\begin{figure*}[t]
    \centering
    \includegraphics[width=\textwidth]{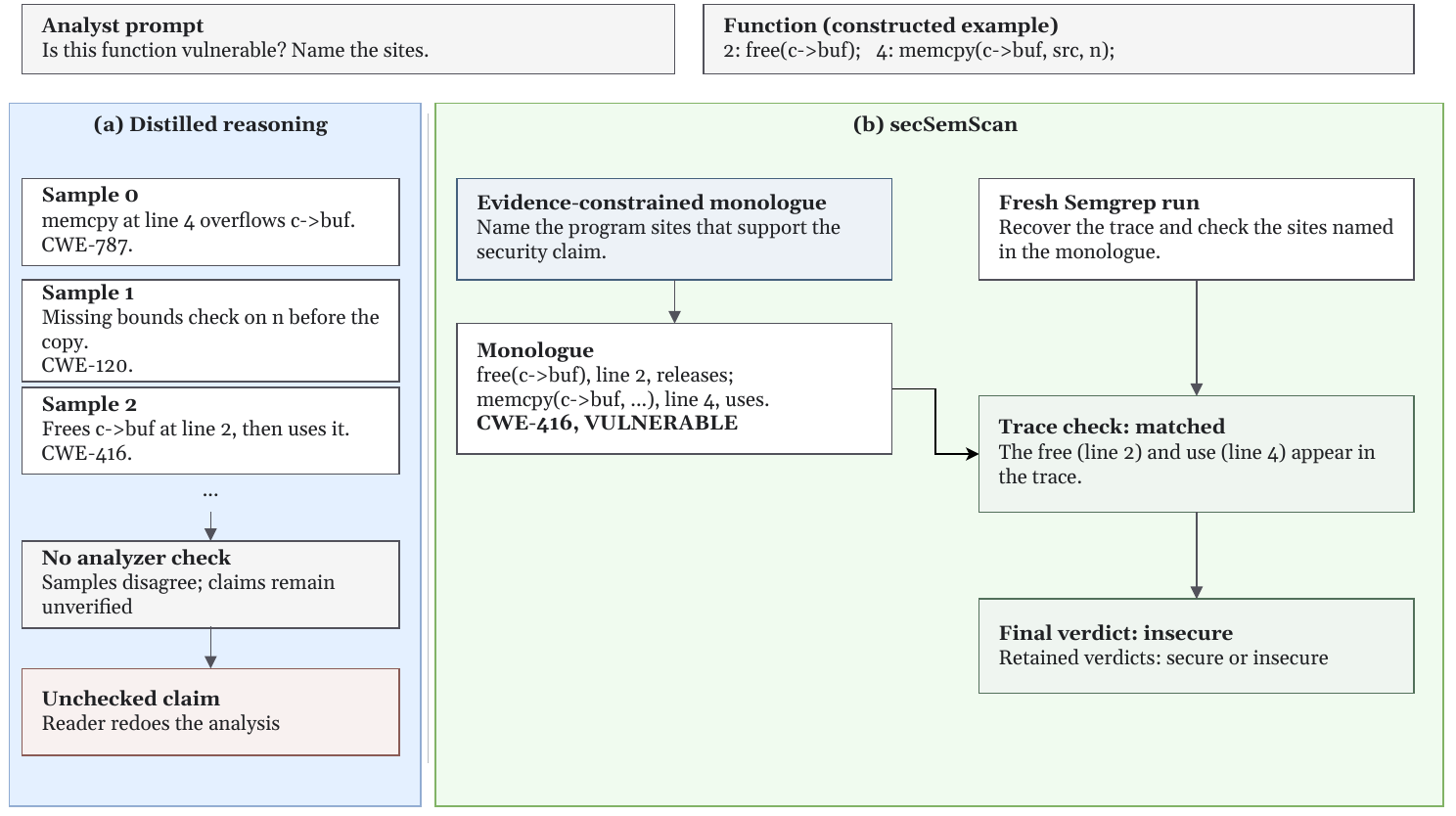}
        \caption{A constructed use-after-free example. Panel (a) shows unchecked free-form reasoning; panel (b) shows secSemScan checking a monologue against a fresh Semgrep run. The illustrated final verdict is \texttt{insecure}; the retained verdicts are \texttt{secure} and \texttt{insecure}. The function and samples are invented for exposition.}

    \label{fig:motivating}
\end{figure*}

Figure~\ref{fig:motivating} illustrates why a security explanation should be grounded in evidence that a tool can check. The analyst asks whether the C function is vulnerable and asks the model to name the relevant sites. At line 2, \texttt{free(c->buf)} releases the buffer; at line 4, \texttt{memcpy(c->buf, src, n)} uses it. Since the pointer is not reset between these operations, the copy accesses freed memory, a use-after-free (CWE-416). These are the same lifetime sites shown in Figure~\ref{fig:monologue_schema}.

Panel (a) presents what can happen when free-form explanations are not checked against the code. The examples are invented to represent the kind of disagreement that can arise in distilled reasoning approaches, i.e., using a teacher LLM. Sample 0 calls the issue a buffer overflow (CWE-787), Sample 1 points to a missing bounds check (CWE-120), and Sample 2 identifies the use-after-free. The reader has no way to tell which account is supported without tracing the code independently. The result is an unchecked claim, which falls short of D1’s requirement that findings be checkable.



Panel (b) shows how secSemScan checks a model-generated monologue. The monologue names the free at line 2 and the use at line 4, and assigns CWE-416. A fresh Semgrep run produces a trace for the function, and secSemScan checks the named sites against it. Since the trace contains both sites, the system returns the \texttt{insecure} verdict shown in the figure. Under our retained binary construction, safe cases receive the \texttt{secure} verdict.

Together, these cases motivate secSemScan's central design: ground model-generated explanations in program sites an analyzer can verify, so that the final secure or insecure verdict rests on checked evidence.

\subsection{Weakness Families and Evidence Tuples}
\label{sec:prelim-weak}

A finding is checkable when it names the program sites that make the code vulnerable. For the two weakness families that secSemScan covers, this evidence is a small tuple of sites, each given by line number and identifier. A \emph{taint weakness}, such as CWE-120, is a flow of user-controlled data from a source to a dangerous sink with no sanitizer on the path, so its evidence is a (source, sink) pair. A \emph{lifetime weakness}, such as CWE-416 or CWE-401, is a violation of the allocate, free, and use sequence of a heap object, so its evidence is an (alloc, free, use) tuple. We call each family a \emph{track}, and Table~\ref{tab:cwe_track_mapping} assigns every CWE in OGSM to one of the two.



\subsection{Security Monologue and Verdict States}
\label{sec:prelim-monologue}

A \emph{security monologue} is a narration, constrained by a fixed schema, that names each site of the weakness by identifier and line, assigns a CWE, and ends with a machine-parseable verdict line (Section~\ref{sec:monologue}). The model writes the monologue, and the oracle checks the sites it names. At inference, secSemScan reconciles the model's verdict with a fresh oracle run on the input function into one of two \emph{verdict states}:

\begin{description}
  \item[\texttt{insecure}] The oracle confirms a path and the monologue reproduces its sites.
  \item[\texttt{secure}] The model reports SAFE and the oracle finds no vulnerable path.
\end{description}

\noindent Section~\ref{sec:method} describes the reconciliation in detail.

%% file: body/data_collection.tex
\section{Dataset}
\label{sec:dataset}

OGSM is a C/C\texttt{++} vulnerability dataset for training and evaluating neuro-symbolic code security analysis. Each record pairs a function with a verdict, a CWE, evidence for the vulnerable site, and a \emph{security monologue}: a schema-constrained narration verified to reproduce the named sites of that evidence.

\subsection{Construction}

\paragraph{Sources.} Candidates come from DiverseVul~\cite{diversevul} (163,776), PairVul~\cite{pairvul} (4,314), and MVD~\cite{mvd} (877), 168,967 functions in total. We exclude functions shorter than 5 or longer than 200 lines. Semgrep confirmed 840 of 41,914 eligible candidates, and 1,215 PairVul candidates had usable \texttt{vuln\_lines} annotations. After selection, merging, and the per-CWE cap, 1,993 records remain (1.2\% of candidates, SAFE records included): 565 from DiverseVul, 1,376 from PairVul, and 52 from MVD. PairVul's own CWE label is reused for about 97\% of its records.



\paragraph{Evidence tiers.} The Semgrep rules cover eight CWE classes, so analyzer-confirmed evidence alone would leave out classes such as CWE-787, CWE-125, CWE-119, and CWE-20. We therefore use two sources of evidence and record the evidence tier for each record. Tier~1 records rest on a static analyzer (Semgrep~\cite{semgrep}), and Tier~2 records rest on the diff of a fix commit. In Tier~1, Semgrep taint and lifetime rules run on each candidate, and a VULNERABLE record needs at least one firing rule. Its trace gives the source and sink, or the alloc, free, and use sites, at line-and-identifier level. Tier~1 has 670 records (335 VULNERABLE, 335 SAFE) from 104 CVEs. In Tier~2, the VULNERABLE or SAFE assignment is PairVul's own pairing of the code before and after the fix, and the evidence is the lines the fix changed, stored as unverified text in the source field (taint) or the alloc field (lifetime). Tier~2 has 1,296 records (648 VULNERABLE, 648 SAFE). Only Tier~1 evidence is analyzer-confirmed; Tier~2 evidence is not checked to be a source, sink, or allocation site. We also tag 27 multi-CWE VULNERABLE records as \texttt{merged}; they carry no trace. Their first-listed CWE is the sole label, is counted in Table~\ref{tab:cwe_track_mapping}, and is checked for CWE accuracy (Section~\ref{sec:setup}) on the one merged test record.


\paragraph{SAFE records.} In Tier~2, the SAFE record is the post-fix function. In Tier~1, 306 of the 335 SAFE records have code identical to their VULNERABLE counterpart with the trace fields removed, and 29 use patched code. For these 306 records, the evidence fields in the prompt carry the label, not the code.

\paragraph{Narrations and capping.} Using Claude Code (Sonnet 4.5) and OpenAI Codex (GPT-5), we generated one narration for each VULNERABLE record based on its code and supporting evidence, retaining all generated narrations without post-generation filtering. SAFE records receive a fixed monologue with verdict \texttt{[SAFE] NO-CWE} and no site narration. To reduce polarity bias toward the most frequent CWEs, we capped VULNERABLE records at 50 per CWE within each split; only the training split required trimming.


\begin{table}
\centering
\caption{CWE-to-track mapping in OGSM (VULNERABLE records, all splits). Merged records are counted under their first-listed CWE.}
\label{tab:cwe_track_mapping}
\begin{tabular}{llrr}
\toprule
CWE & Description & Track & VULN records \\
\midrule
CWE-20  & Improper input validation         & Taint & 103 \\
CWE-362 & Race condition (TOCTOU)           & Taint & 100 \\
CWE-119 & Buffer errors                     & Taint &  81 \\
CWE-476 & NULL pointer dereference          & Taint &  78 \\
CWE-787 & Out-of-bounds write               & Taint &  77 \\
CWE-190 & Integer overflow                  & Taint &  74 \\
CWE-125 & Out-of-bounds read                & Taint &  74 \\
CWE-400 & Uncontrolled resource consumption & Taint &  62 \\
CWE-667 & Improper locking                  & Taint &  32 \\
CWE-835 & Infinite loop                     & Taint &  27 \\
CWE-367 & TOCTOU race (file)                & Taint &  22 \\
CWE-120 & Classic buffer overflow           & Taint &  19 \\
CWE-843 & Type confusion                    & Taint &  12 \\
CWE-755 & Improper exception handling       & Taint &   9 \\
CWE-665 & Improper initialization           & Taint &   9 \\
\midrule
CWE-401 & Missing free                      & Lifetime & 90 \\
CWE-416 & Use-after-free                    & Lifetime & 66 \\
CWE-459 & Incomplete cleanup                & Lifetime & 22 \\
CWE-770 & Allocation without limits         & Lifetime & 17 \\
CWE-404 & Improper resource shutdown        & Lifetime & 14 \\
CWE-908 & Use of uninitialized resource     & Lifetime & 13 \\
CWE-415 & Double-free                       & Lifetime &  9 \\
\midrule
\multicolumn{3}{r}{Total} & 1,010 \\
\bottomrule
\end{tabular}
\end{table}

\paragraph{Tracks.} Each CWE is assigned to the taint or lifetime track by its defining mechanism, recorded in the \texttt{task} field (Table~\ref{tab:cwe_track_mapping}). We place CWE-476 in the taint track because a null pointer is the sink of an unchecked flow, and CWE-362 because the check-to-use window is the unsanitized path of the race. The cap applies within each split while the table sums over splits, so a total can exceed 50: CWE-20 (103) and CWE-362 (100) each have at most 50 per split.

\subsection{Statistics and Splits}
 
\begin{table}
\centering
\caption{OGSM dataset statistics.}
\label{tab:dataset_stats}
\begin{tabular}{lrrrr}
\toprule
Split & Records & VULNERABLE & SAFE & Unique CWEs \\
\midrule
Train & 1,248 & 637 & 611 & 22 \\
Val   &   390 & 195 & 195 & 16 \\
Test  &   355 & 178 & 177 & 22 \\
\midrule
Total & 1,993 & 1,010 & 983 & 22 \\
\bottomrule
\end{tabular}
\end{table}

OGSM comprises 1,993 records spanning 22 CWEs and 433 CVEs (Table~\ref{tab:dataset_stats}). The taint track contains 1,531 records (779 VULNERABLE, 752 SAFE), while the lifetime track contains 462 (231 each). The 178 VULNERABLE test records comprise 55 Tier~1, 122 Tier~2, and one merged record, sourced from DiverseVul (47), PairVul (127), and MVD (4). Data splits are CVE-disjoint, ensuring that no CVE appears in multiple splits. PairVul contributes 1,376 of the 1,993 records and 127 of the 178 VULNERABLE test records, with records represented in both training and test sets. Since Tier~2 labels derive from PairVul's own pairing, Tier~2 results measure agreement with these labels and may partly reflect PairVul-specific patterns (Section~\ref{sec:eval}).


\subsection{Comparison with Related Datasets}
 
Table~\ref{tab:dataset_comparison} compares OGSM with common vulnerability datasets. All of them give function-level labels with CVE and CWE identifiers, and several include the fix patch. OGSM has 1,993 records, against 5.8K to 349K for the others. It differs in its Semgrep trace, present in Tier~1 records only, and in its narrations, which are LLM-generated and unfiltered.
 
\begin{table}
\centering
\caption{Comparison of OGSM and related datasets.}
\label{tab:dataset_comparison}
\begin{tabular}{lrrcccc}
\toprule
Dataset & Records & Vuln. funcs & CVE/CWE & Fix patch & Analyzer trace & Narration \\
\midrule
DiverseVul~\cite{diversevul} & 349K  & 18,945 & \cmark & \xmark & \xmark & \xmark \\
PairVul~\cite{pairvul}       &  5.8K &  2,903 & \cmark & \cmark & \xmark & \xmark \\
MVD~\cite{mvd}               &  8.7K &  4,353 & \cmark & \cmark & \xmark & \xmark \\
\midrule
\textbf{OGSM (ours)} & \textbf{1,993} & \textbf{1,010} & \cmark & \xmark & \cmark  & \cmark \\
\bottomrule
\end{tabular}
\end{table}

%% file: body/study_design.tex
\begin{figure}
    \centering
    \includegraphics[width=0.9\linewidth]{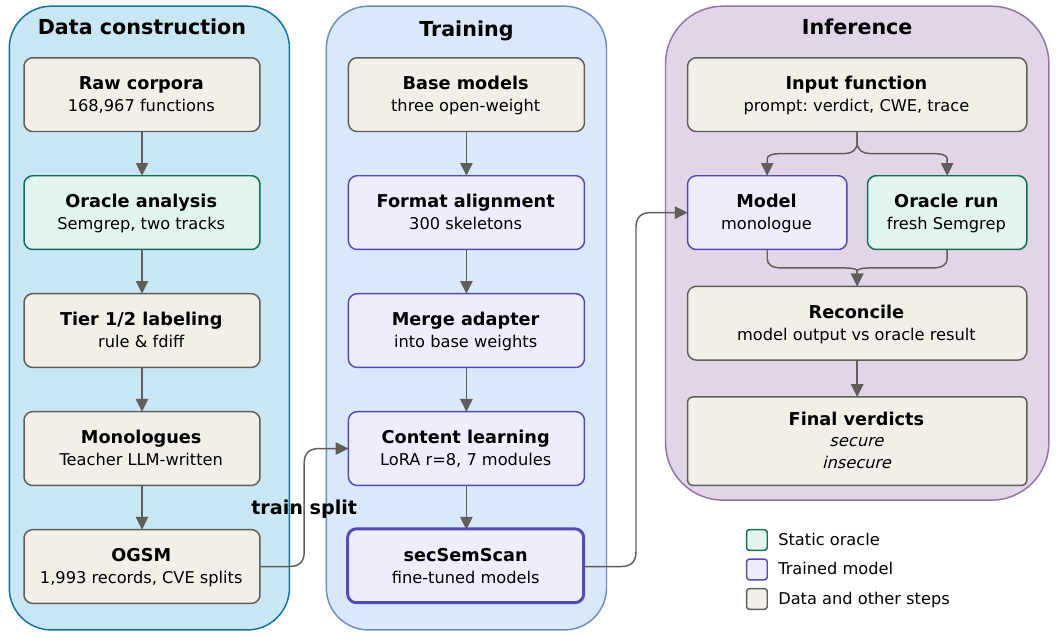}
    \caption{The secSemScan pipeline. Oracle traces and diff-derived evidence are converted into security monologues (the OGSM dataset), which train the model in a two-stage curriculum. At inference, the model's output is reconciled with a fresh oracle run into one of two verdicts: \texttt{secure} or \texttt{insecure}.}
\label{fig:methodology_overview}
\end{figure}

\section{Methodology}
\label{sec:method}

secSemScan fine-tunes an open-weight code model to write a security monologue that names the program sites of a vulnerability, and it checks the model's verdict against a static analyzer at inference. Figure~\ref{fig:methodology_overview} shows the five stages: data collection, oracle analysis, training-data construction, training, and inference. Section~\ref{sec:dataset} describes data collection, the evidence tiers, narration generation, and the splits. This section describes the oracle, the monologue and prompt design, the training procedure, and inference.

\subsection{Oracle Analysis}
 
We use Semgrep~\cite{semgrep} as the oracle and run two rule tracks on every function. The \emph{taint track} checks for dataflow from user-controlled sources, such as \texttt{copy\_from\_user} and ioctl or netlink entry points, to dangerous sinks, such as \texttt{memcpy} size arguments, array indices, and format strings, and reports a finding only when no sanitizer lies on the path. The \emph{lifetime track} checks the alloc, free, and use sequence of heap objects and reports use-after-free (CWE-416), double-free (CWE-415), and missing-free (CWE-401), among others. Each track yields a trace: a tuple of named sites at line-and-identifier level. A firing rule gives a Tier~1 record its VULNERABLE label and trace; Tier~2 records rest on the diff of a fix commit instead (Section~\ref{sec:dataset}). Because the same rules label Tier~1 records and run at inference, agreement with the oracle on Tier~1 is not independent evidence (Section~\ref{sec:threats}).



\subsection{Monologue and Prompt Design}
\label{sec:monologue}

An oracle trace is a tuple of program sites. Although tools can check it, developers may struggle to triage it, while free-form prose lacks machine-checkable structure (Section~\ref{sec:intro}). A \emph{security monologue} follows a fixed schema, identifying vulnerability sites by identifier and line, assigning the CWE, and ending with a machine-parseable verdict. Unlike SemCoder~\cite{ding2024semcoder}, which grounds narrations in execution traces, we use the static oracle's trace because taint propagation and object lifetime are static program facts (Section~\ref{sec:background}). Section~\ref{sec:dataset} details narration generation.

\begin{figure}
    \centering
    \includegraphics[width=\linewidth]{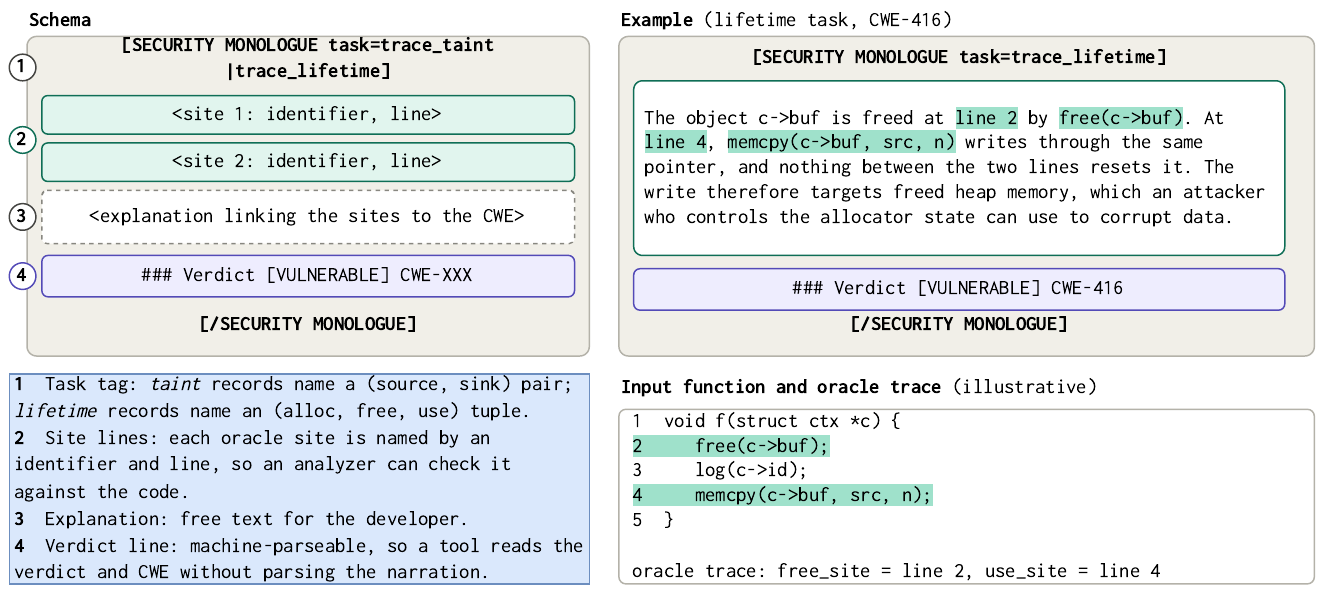}
    \caption{The security monologue schema (left) and an illustrative lifetime example (right). The narration names the oracle's free and use sites by identifier and line, and the last line gives a machine-parseable verdict and CWE. The example is constructed for exposition and is not an OGSM record.}
    \label{fig:monologue_schema}
\end{figure}

The schema comprises four parts (Figure~\ref{fig:monologue_schema}, left). An opening tag specifies \texttt{trace\_taint} or \texttt{trace\_lifetime}, determining the required sites: source and sink for taint, or allocation, free, and use for lifetime. The narration identifies these sites by identifier and line, then briefly explains how they violate the CWE. A fixed-format verdict line enables machine parsing without interpreting the explanation, and a closing tag terminates the monologue. 

The right side of Figure~\ref{fig:monologue_schema} illustrates a constructed use-after-free in a five-line function. The oracle identifies \texttt{free(c->buf)} at line 2 and \texttt{memcpy(c->buf, src, n)} at line 4. The narration names both sites, notes that the pointer is not reset between them, and explains the exploitable write before assigning CWE-416. Only two lifetime sites appear because the function contains no allocation; this example is illustrative, not an OGSM record. SAFE records instead receive a fixed monologue containing \texttt{[SAFE] NO-CWE}, without site narration. All training and evaluation prompts use a fixed template specifying the function, CWE, oracle mode, trace, and verdict, and requesting a narration of at most 150 words in the prescribed schema. Thus, the model narrates a supplied trace, which defines the scope of evaluation (Section~\ref{sec:eval}).

\subsection{Model Training}
\label{sec:training}
 

We fine-tune Qwen2.5-Coder-7B~\cite{hui2024qwen25coder}, DeepSeek-Coder-6.7B~\cite{guo2024deepseekcoder}, and StarCoder2-7B~\cite{lozhkov2024starcoder2} with supervised fine-tuning and low-rank adaptation (LoRA)~\cite{hu2022lora}, using the same two-stage curriculum for all three. The curriculum separates learning the schema from learning its content. We adopt it as a design choice and do not compare it with single-stage training (Section~\ref{sec:threats}). Each training example pairs a prompt $x$ (Section~\ref{sec:monologue}) with a target monologue $y = (y_1, \dots, y_T)$. The target is the stored narration for a VULNERABLE record and the fixed monologue for a SAFE record. The model minimizes the negative log-likelihood of the target tokens given the prompt, and both stages mask the prompt tokens so that only the output contributes to the loss. LoRA freezes a pretrained weight matrix $W \in \mathbb{R}^{d \times k}$ and learns a low-rank update $BA$, where $B \in \mathbb{R}^{d \times r}$ and $A \in \mathbb{R}^{r \times k}$ have rank $r \ll \min(d, k)$. 


\textbf{Stage 1: format alignment.} The base models have not seen the monologue schema, so the first stage teaches only the format. We train on 300 synthetic skeletons instantiated from five templates, three VULNERABLE and two SAFE, which cover both task types. Each skeleton is a correctly formatted monologue with a placeholder narration, so the stage carries no CWE-specific content. Training continues until format compliance (FC, Section~\ref{sec:setup}) exceeds 90\% on a held-out format check set. We then merge the resulting adapter into the base weights, $W_1 = W_0 + B_1 A_1$.

\textbf{Stage 2: content learning.} We initialize a fresh adapter $(B_2, A_2)$ on $W_1$ and train it on the 1,248 OGSM training records (637 VULNERABLE, 611 SAFE) with the same masked loss. Merging before this stage keeps the format-aligned weights fixed while the new adapter learns the content, and the ablation tests whether this ordering matters.
 


 
 

\subsection{Inference}

At inference, secSemScan prompts the fine-tuned model for a monologue and reads the verdict from its last line: \texttt{insecure} if the monologue reports VULNERABLE, and \texttt{secure} if it reports SAFE. An output without a verdict line has no verdict, and we count it as wrong in the evaluation (Section~\ref{sec:setup}). For an \texttt{insecure} verdict, a fresh Semgrep run on the input function checks the sites the monologue names, so a developer or a tool can confirm the finding without repeating the analysis. A \texttt{secure} verdict is the model's claim alone, because a rule-based analyzer that stays silent does not show that a function is safe (Section~\ref{sec:background}). In our evaluation the prompt also states the record's verdict, CWE, and oracle trace, so the evaluation measures whether the model turns a given trace into a schema-conforming narration, not whether it finds vulnerabilities in code alone.

%% file: body/resullt.tex
\section{Evaluation}
\label{sec:eval}

\subsection{Experimental Setup}
\label{sec:setup}

\paragraph{Test set.} The test set contains 355 OGSM records, 178 VULNERABLE and 177 SAFE, covering 22 CWEs (Table~\ref{tab:dataset_stats}). The VULNERABLE records comprise 55 Tier~1 records, 122 Tier~2 records, and one merged record. The splits are CVE-disjoint, but PairVul appears in both training and test and contributes 127 of the 178 VULNERABLE test records (Section~\ref{sec:dataset}). Every prompt states the record's verdict, CWE, oracle mode, and oracle evidence (Section~\ref{sec:monologue}), so the evaluation measures whether a model turns supplied evidence into a schema-conforming narration, not whether it detects vulnerabilities in code alone.

\paragraph{Models and baselines.} We fine-tune three open-weight base models, Qwen2.5-Coder-7B, DeepSeek-Coder-6.7B, and StarCoder2-7B, with the same two-stage curriculum (Section~\ref{sec:training}). Qwen2.5-Coder-7B is the headline model because it has the highest verdict and CWE accuracy (Table~\ref{tab:overall}). The baseline for each is the untuned model with the same prompt, schema, and decoding. We score untuned models with a strict parser that accepts only the monologue schema and a lenient parser that takes the first verdict keyword and CWE in free text; secSemScan is scored with the strict parser. A record without a parsable verdict counts as wrong.

\paragraph{Implementation.} We use Hugging Face \texttt{transformers} with bf16 weights, \texttt{peft}, and \texttt{trl} on NVIDIA L40S GPUs, with one GPU per evaluation run. Decoding is greedy with at most 512 new tokens, and inputs are truncated at 4,096 tokens. The sampling experiment (Section~\ref{sec:robustness}) uses its own settings. Training scripts and hyperparameters are in the artifact. 


\subsection{Performance of the Finetuned Models}
\label{sec:overall}

In this subsection, we analyze whether fine-tuning on oracle-verified monologues improves verdict and CWE prediction over the untuned base model, and whether the improvement holds across base models. We compare secSemScan, the fine-tuned variant, with the untuned base model for Qwen2.5-Coder-7B, DeepSeek-Coder-6.7B, and StarCoder2-7B (Table~\ref{tab:overall}). We then break down the outcomes of secSemScan by label and examine its errors (Table~\ref{tab:breakdown}), and finally report recall by evidence tier. The prompt states the record's verdict and CWE and supplies the oracle evidence (Section~\ref{sec:monologue}), so these results measure whether a model turns a supplied trace into a narration with the right verdict and CWE. 

All models are evaluated on the 355 test records (178 VULNERABLE, 177 SAFE) with greedy decoding. Verdict accuracy (VA) is the share of records whose parsed verdict matches the label, and CWE accuracy is the share of the 178 VULNERABLE records whose predicted CWE matches the label. A record with no parsable verdict counts as wrong for both. Format compliance (FC), the share of outputs that follow the full monologue schema, is secondary. We score each base model with two parsers. The strict parser accepts only the monologue schema, and the lenient parser takes the first verdict keyword and the first CWE found in free text, so it shows what the base model produces apart from format. secSemScan is scored with the strict parser.

\begin{table}[t]
\centering
\caption{Overall performance on the 355 test records. CWE accuracy uses the 178 VULNERABLE records. FC does not apply to the lenient parser.}
\label{tab:overall}
\begin{tabular}{llrrr}
\toprule
Base model & Variant & VA (\%) & CWE (\%) & FC (\%) \\
\midrule
Qwen2.5-Coder-7B    & base, strict  &  4.8 &  5.1 &  4.8 \\
                    & base, lenient & 41.7 & 46.1 & --   \\
                    & secSemScan    & 96.6 & 96.6 & 96.6 \\
\midrule
DeepSeek-Coder-6.7B & base, strict  &  1.1 &  1.7 &  0.0 \\
                    & base, lenient &  1.1 &  1.7 & --   \\
                    & secSemScan    & 78.9 & 71.3 &  0.0 \\
\midrule
StarCoder2-7B       & base, strict  &  5.9 &  1.1 &  5.1 \\
                    & base, lenient &  6.8 &  1.7 & --   \\
                    & secSemScan    & 43.7 & 51.7 & 18.0 \\
\bottomrule
\end{tabular}
\end{table}

On Qwen2.5-Coder-7B, fine-tuning raised VA from 4.8\% to 96.6\% (343 of 355) and CWE accuracy from 5.1\% to 96.6\% (172 of 178) (Table~\ref{tab:overall}). The lenient parser raised the base model's VA only to 41.7\% (148 of 355). It extracted a verdict from 150 base outputs and 148 of them were correct, while the other 205 outputs contain no verdict keyword. Because the prompt states the verdict, a correct verdict is the expected outcome whenever one appears.
 
\observation{Fine-tuning raised the VA and CWE accuracy of Qwen2.5-Coder-7B from 4.8\% and 5.1\% to 96.6\% each. The lenient parser raised the base model's VA only to 41.7\%, and its misses are outputs without a verdict and not wrong verdicts.}
 
On the other two base models, fine-tuning also improved VA and CWE accuracy, but by less than on Qwen2.5-Coder-7B. VA rose from 1.1\% to 78.9\% for DeepSeek-Coder-6.7B and from 5.9\% to 43.7\% for StarCoder2-7B, and CWE accuracy rose from 1.7\% to 71.3\% and from 1.1\% to 51.7\%. No output of DeepSeek-Coder-6.7B satisfied the full schema, so its FC is 0.0\%.
 
\observation{The gain from fine-tuning depends on the base model. VA after fine-tuning is 96.6\% for Qwen2.5-Coder-7B, 78.9\% for DeepSeek-Coder-6.7B, and 43.7\% for StarCoder2-7B.}

\begin{table}[t]
\centering
\caption{Breakdown of secSemScan outcomes by label on the 355 test records.}
\label{tab:breakdown}
\begin{tabular}{lrrrr}
\toprule
Base model & VULNERABLE (178) & SAFE (177) & Wrong verdict & No verdict \\
\midrule
Qwen2.5-Coder-7B    & 172 & 171 & 0 &  12 \\
DeepSeek-Coder-6.7B & 130 & 150 & 0 &  75 \\
StarCoder2-7B       &  93 &  62 & 0 & 200 \\
\bottomrule
\end{tabular}
\end{table}

We analyzed the errors of secSemScan by label and by input function (Table~\ref{tab:breakdown}). None of the three models gave a wrong verdict, so every error is an output without a verdict. The 12 such outputs on Qwen2.5-Coder-7B come from six functions, and each function fails under both its VULNERABLE and its SAFE label. The failure therefore depends on the function and not on the label. Four of the six functions are the four MVD functions in the test set, and the other two are one DiverseVul function and one PairVul function.
 
\observation{Every error of the tuned models is an output without a verdict. On Qwen2.5-Coder-7B, the 12 errors come from six functions that each fail under both labels, and they include all four MVD functions in the test set.}
 
Among the VULNERABLE records, Qwen2.5-Coder-7B secSemScan returned the correct verdict on 50 of 55 Tier~1 records (90.9\%) and on 121 of 122 Tier~2 records (99.2\%). By construction, Semgrep alone fires on all 55 Tier~1 records and on none of the 122 Tier~2 records. The Tier~2 figure is limited in two ways: the prompt states the verdict, and Tier~2 labels come from PairVul, which also appears in training (Section~\ref{sec:dataset}).
 
\observation{Qwen2.5-Coder-7B secSemScan returned the correct verdict on 90.9\% of Tier~1 and 99.2\% of Tier~2 VULNERABLE records, while Semgrep alone fires on all Tier~1 records and on none of the Tier~2 records by construction. The prompt states the verdict, so the Tier~2 figure does not show that the model finds vulnerabilities that Semgrep misses.}


\subsection{Grounding of Monologues in the Oracle Evidence}
\label{sec:quality}

This subsection asks whether the narrations that secSemScan produces name the sites the oracle identified, stay grounded in the function code, and agree with their own verdict. The prompt lists the oracle sites (Section~\ref{sec:monologue}), so these checks measure whether a narration uses the supplied evidence, not whether it finds the sites unaided. We apply eight rule-based checks to the cached outputs of the untuned and fine-tuned models, using no judge model, so every rate is a deterministic count over the outputs (Table~\ref{tab:quality}). Header present is the share of the 355 outputs that begin with the monologue header, and mode matches track is the share of those headers whose mode (taint or lifetime) equals the record's track. Tier~1 sites cited is the share of the 80 oracle sites in the 55 Tier~1 VULNERABLE records whose line number appears in the output, and cited lines that are oracle sites is the share of cited line numbers that belong to the oracle. Tier~2 evidence quoted is the share of the 122 Tier~2 records whose output contains the evidence text from the PairVul diff. Identifiers found in code is the share of backticked identifiers that occur verbatim in the function. CWE in body matches verdict is the share of VULNERABLE outputs with a verdict whose body names the same CWE as the verdict line. Repeated sentences is the share of sentences that duplicate an earlier sentence in the same output.

\begin{table}[t]
\centering
\caption{Rule-based checks on monologue quality (percent; a fraction replaces the rate when the denominator is below 10). Repeated sentences: lower is better. A dash means the check does not apply or is not reported.}
\label{tab:quality}
\resizebox{\columnwidth}{!}{%
\begin{tabular}{lrrrrr}
\toprule
 & \multicolumn{2}{c}{Qwen2.5-Coder-7B} & \multicolumn{2}{c}{StarCoder2-7B} & DeepSeek-6.7B \\
\cmidrule(lr){2-3}\cmidrule(lr){4-5}\cmidrule(lr){6-6}
Check & base & secSemScan & base & secSemScan & secSemScan \\
\midrule
Header present                      & 13.2 & 96.6  &  0.0 & 19.2 & --   \\
Mode matches track                  & --   & 100.0 & --   & 100.0 & --  \\
Tier 1 sites cited                  & 60.0 & 87.5  &  5.0 & 53.8 & 76.2 \\
Cited lines that are oracle sites   & 96.0 & 100.0 & 4/4  & 97.7 & 98.4 \\
Tier 2 evidence quoted              & 57.4 & 57.4  &  1.6 & 72.1 & 32.8 \\
Identifiers found in code           & 94.6 & 94.6  & --   & 93.7 & 84.3 \\
CWE in body matches verdict         & 5/9  & 95.9  & 1/2  & 95.7 & 91.5 \\
Repeated sentences                  &  5.4 &  0.0  & 22.9 & 21.8 & --   \\
\bottomrule
\end{tabular}}
\end{table}

On Qwen2.5-Coder-7B, fine-tuning raised the share of outputs that begin with the header from 13.2\% (47 of 355) to 96.6\% (343 of 355), and the mode in all 343 headers matches the record's track. The fine-tuned model cites 70 of the 80 Tier~1 sites (87.5\%), and every line it cites is an oracle site (70 of 70). The untuned model cites 48 of 80 (60.0\%), so fine-tuning raises site citation by 27.5 points. On the 177 SAFE records, no system cites a line that is absent from the evidence.
 
\observation{On Qwen2.5-Coder-7B, secSemScan cites 87.5\% of the Tier~1 oracle sites (base 60.0\%) and never cites a line outside the oracle. The prompt supplies the sites, so this measures use of the evidence and not discovery.}

For Tier~2 records the evidence is code text from the PairVul diff, without line numbers. Qwen2.5-Coder-7B secSemScan quotes it in 70 of 122 records (57.4\%), the same count as the untuned model, although the two models quote it on different records (50 records in common). Both models also name identifiers that occur in the function at the same rate: 1,868 of 1,974 (94.6\%) for secSemScan and 2,225 of 2,353 (94.6\%) for the untuned model.
 
\observation{Fine-tuning does not change how often the narration quotes the Tier~2 evidence (57.4\% for both) or how often the identifiers it names occur in the code (94.6\% for both). The gains on Qwen2.5-Coder-7B are in format and in citing the Tier~1 sites.}

Among the 172 VULNERABLE outputs of Qwen2.5-Coder-7B secSemScan that carry a verdict, 165 (95.9\%) name in the body the same CWE as the verdict line, and in 6 of the other 7 the body names no CWE. No sentence of its 2,051 repeats an earlier sentence in the same output, against 139 of 2,580 (5.4\%) for the untuned model.
 
The other two models score lower on most checks. StarCoder2-7B secSemScan begins with the header in 68 of 355 outputs (19.2\%), cites 43 of 80 Tier~1 sites (53.8\%), and repeats 21.8\% of its sentences (566 of 2,593), with 19 outputs in which more than half the sentences repeat. It quotes the Tier~2 evidence more often than Qwen2.5-Coder-7B does (72.1\%). DeepSeek-Coder-6.7B secSemScan cites 61 of 80 Tier~1 sites (76.2\%) with 61 of 62 cited lines correct, but quotes the Tier~2 evidence in 40 of 122 records (32.8\%), and 84.3\% of its identifiers occur in the code.
 
\observation{Fine-tuning produced well-formed, non-repetitive narrations that cite the Tier~1 evidence on Qwen2.5-Coder-7B (header 96.6\%, citation 87.5\%, repeated sentences 0.0\%). StarCoder2-7B secSemScan cites 53.8\% of the sites, and 21.8\% of its sentences repeat.}

These checks test the surface of a narration, not whether its explanation is correct. We read 14 narrations drawn with a fixed seed from four groups (Tier~1 VULNERABLE, Tier~2 VULNERABLE, SAFE, and outputs without a verdict). In a Tier~1 sample (\texttt{DiverseVul\_nocve\_28936}), the narration cites both oracle sites, the allocation at line 18 and the dereference at line 37, and states CWE-476 with a mechanism that follows from them. In a Tier~2 sample (\texttt{vl\_CVE-2013-2146\_101}), the narration quotes the evidence and gives the correct CWE-20, but its description of the flow assigns \texttt{x86\_pmu.extra\_regs} to itself and does not explain the flaw. Both narrations would pass the checks in Table~\ref{tab:quality}, so a grounded narration is not necessarily a correct one, and we did not review the explanations against a reference.


\subsection{Robustness to Decoding Choices}
\label{sec:robustness}

In this experiment, we analyze whether the verdicts of secSemScan depend on greedy decoding, whether the 12 Qwen2.5-Coder-7B records without a verdict under greedy decoding recover under sampling, and whether the effect of fine-tuning holds across the three base models. The prompt states the verdict, CWE, and oracle trace (Section~\ref{sec:monologue}), so the experiment tests the stability of the narration given the answer, not the stability of detection.
 
We drew $n=5$ samples for each of the 355 test records, for each untuned base model and its secSemScan variant, with temperature 0.8, top-$p$ 0.95, and seed 20261002. All other settings are identical for the six systems and follow the main evaluation, including the strict parser, so a sample without a verdict line counts as wrong. We report pass@$k$, the probability that at least one of $k$ samples drawn from the $n$ has a correct verdict, computed with the unbiased estimator. Pass@1 is the expected accuracy of a single output and is comparable to the greedy VA. Pass@3 and pass@5 measure headroom, not the accuracy of a system that returns one output.

\begin{table}[t]
\centering
\caption{Verdict pass@$k$ (\%) over the 355 test records ($n=5$, temperature 0.8, top-$p$ 0.95). }
\label{tab:passk}
\begin{tabular}{llrrrr}
\toprule
Base model & Variant & Greedy VA & pass@1 & pass@3 & pass@5 \\
\midrule
Qwen2.5-Coder-7B    & base       &  4.8 & 16.7 [13.1, 20.8] & 41.8 [36.7, 46.9] & 58.9 [53.7, 63.9] \\
                    & secSemScan & 96.6 & 96.6 [94.2, 98.1] & 96.6 [94.2, 98.1] & 96.6 [94.2, 98.1] \\
\midrule
DeepSeek-Coder-6.7B & base       &  1.1 & 31.6 [23.8, 40.5] & 66.0 [57.0, 74.0] & 82.8 [74.9, 88.6 \\
                    & secSemScan & 78.9 & 87.2 [83.3, 90.3] & 95.8 [93.1, 97.4] & 96.1 [93.5, 97.6] \\
\midrule
StarCoder2-7B       & base       &  5.9 & 24.0 [19.9, 28.7] & 53.9 [48.7, 59.0] & 70.7 [65.8, 75.2] \\
                    & secSemScan & 43.7 & 37.5 [32.6, 42.6] & 74.5 [69.7, 78.8] & 89.3 [85.6, 92.1] \\
\bottomrule
\end{tabular}
\end{table}
 
For Qwen2.5-Coder-7B secSemScan, pass@1, pass@3, and pass@5 are all 96.6\% (343 of 355) (Table~\ref{tab:passk}), equal to the greedy VA. Every record receives the same outcome in all five samples. The 12 records without a verdict are six functions with two records each, matching the six over-length prompts of the main evaluation (Section~\ref{sec:overall}), and none produces a parsable verdict in any sample. The untuned model differs: its pass@1 is 16.7\% and its pass@5 is 58.9\% (209 of 355), so its output depends on the sample.
 
\observation{Sampling does not change the VA of Qwen2.5-Coder-7B secSemScan (96.6\% at every $k$), and the 12 records without a verdict under greedy decoding have none in any of five samples. The untuned model reaches 16.7\% pass@1 and 58.9\% pass@5.}

 The other two fine-tuned models vary across samples. StarCoder2-7B secSemScan reaches 37.5\% pass@1, below its greedy VA of 43.7\%, and 89.3\% pass@5, so its verdict depends on the sample, although a deployed system returns one output. DeepSeek-Coder-6.7B secSemScan reaches 87.2\% pass@1, below its greedy VA of 78.9\%, and 96.1\% pass@5. For all three untuned models, pass@1 exceeds the greedy VA (16.7\% against 4.8\% for Qwen2.5-Coder-7B, 31.6\% against 1.1\% for DeepSeek-Coder-6.7B, and 24.0\% against 5.9\% for StarCoder2-7B), and the untuned DeepSeek-Coder-6.7B reaches 82.8\% pass@5 although it returned no verdict under greedy decoding.


\observation{The effect of fine-tuning holds under sampling on all three base models: pass@1 rises from 16.7\% to 96.6\% on Qwen2.5-Coder-7B (79.9 points), from 31.6\% to 87.2\% on DeepSeek-Coder-6.7B (55.6 points), and from 24.0\% to 37.5\% on StarCoder2-7B (13.5 points), and the fine-tuned model is higher at pass@3 and pass@5 as well. The size of the gain depends on the base model. Pass@1 of the fine-tuned DeepSeek-Coder-6.7B and StarCoder2-7B is below the greedy VA of each, so sampling lowers the accuracy of a single output on these two models.}

\subsection{Case Study: Performance on the Linux Kernel Dataset}
\label{sec:casestudy}

The benchmark in Section~\ref{sec:eval} measures accuracy on held-out records drawn from the same source corpora as training, so it cannot answer a different question: does secSemScan detect real vulnerabilities in code entirely outside that distribution, in cases where the static oracle also provides no confirming signal of its own? This case study answers that question directly, using kernel security fix commits as ground truth rather than any label inherited from the OGSM construction pipeline.

\subsubsection{Protocol}

We take five functions that later security fix commits changed. For each function we use the version from the parent of its fix commit, and secSemScan runs on it with no fix information. We also run Semgrep alone on the same functions. The ground truth for each function is the fix commit of its CVE. The CWE label comes from the NVD entry, confirmed against the commit message. We check every CVE identifier of the ground truth against \texttt{train.jsonl} and \texttt{val.jsonl}, and drop any overlap. The two CVEs that overlap with OGSM, CVE-2018-8897 and CVE-2020-29374, affect other functions and are excluded.

\subsubsection{Results}

secSemScan labels all 5 functions VULNERABLE with the correct CWE (Table~\ref{tab:casestudy}). None of the 5 CVEs appears in any OGSM split, and Semgrep fires no rule on any of the 5 functions before the fix -- so on this set, the oracle that verifies secSemScan's training data cannot itself confirm the model's predictions, and the fix commits are the only available ground truth.

\begin{table}[h]
\centering
\caption{Five CVE-2024 kernel functions. All five verdicts fall in the \texttt{unverified\_trust} state.}
\label{tab:casestudy}
\resizebox{\columnwidth}{!}{%
\begin{tabular}{llllr}
\toprule
CVE & Function & File & CWE & Correct \\
\midrule
CVE-2024-50047 & \texttt{crypt\_message}            & fs/smb/client/smb2ops.c                  & CWE-416 & \cmark \\
CVE-2024-25741 & \texttt{printer\_read}             & drivers/usb/gadget/function/f\_printer.c & CWE-416 & \cmark \\
CVE-2024-41090 & \texttt{tap\_get\_user\_xdp}       & drivers/net/tap.c                        & CWE-416 & \cmark \\
CVE-2024-24858 & \texttt{conn\_info\_min\_age\_set} & net/bluetooth/hci\_debugfs.c             & CWE-362 & \cmark \\
CVE-2024-46701 & \texttt{simple\_offset\_add}       & fs/libfs.c                               & CWE-190 & \cmark \\
\midrule
\multicolumn{4}{r}{True positive rate} & 5 of 5 \\
\bottomrule
\end{tabular}}
\end{table}

\observation{secSemScan flags all 5 kernel functions as VULNERABLE with the correct CWE. With only 5 functions, the 95\% interval for the rate runs from about 57\% to 100\%; the result shows the method can work on unseen, oracle-silent code but does not give a reliable rate.}

\observation{All 5 verdicts fall in the \texttt{unverified\_trust} state, since Semgrep is silent on each function. The fix commits confirm claims that only the model made, which is the specific gap in benchmark coverage this case study is meant to probe.}

We show two examples that cover both oracle tracks, taint and lifetime, and whose monologues name the exact identifier that the fix commit changes. For CVE-2024-24858, secSemScan writes: \emph{``the \texttt{hdev} pointer is obtained from \texttt{debugfs}'s \texttt{private\_data} without acquiring the lock (\texttt{hci\_dev\_lock}). This makes the \texttt{hdev} pointer tainted, as it may reference a device that is being freed concurrently via \texttt{hci\_unregister\_dev}. The tainted \texttt{hdev} is then used to write to \texttt{hdev->conn\_info\_min\_age} after the lock is released.''} The fix commit \texttt{7835fcfd132e} adds the missing lock. For CVE-2024-46701, secSemScan writes: \emph{``\texttt{octx->next\_offset} is incremented in \texttt{mtree\_alloc\_cyclic} without an overflow check. If this offset exceeds \texttt{LONG\_MAX}, it may wrap around to 0, corrupting the directory read position.''} The fix commit \texttt{a0634b457eca} adds an overflow guard.

%% file: body/threads.tex
\section{Threats to validity}
\label{sec:threats}

\paragraph{Construct validity.} We operationalize monologue usefulness through checkable grounding and consistency properties, including oracle-site citation, identifier grounding, and CWE consistency. These rule-based checks avoid an LLM judge but do not directly assess explanation quality or developer utility. Because the prompt supplies the verdict, CWE, and oracle evidence, our evaluation concerns evidence-grounded narration rather than vulnerability detection.

\paragraph{Internal validity.} We compare each fine-tuned model with its untuned base under identical prompts, schemas, and decoding settings. We report evidence tiers separately because Tier~1 uses Semgrep-derived evidence, while Tier~2 may retain dataset-specific characteristics from PairVul despite CVE-disjoint splits. LLM-generated training narrations may also introduce noise.

\paragraph{External validity.} OGSM focuses on C/C\texttt{++} and two weakness tracks, limiting generalization to other languages and vulnerability classes. We additionally evaluate on held-out kernel vulnerabilities without training/validation CVE overlap, although the case study is small. All experiments use the same software and hardware environment; we provide scripts and configurations to support replication.

\paragraph{Statistical validity.} We observe improvements over the corresponding base model on all three base models, and the sampling experiment (five samples per record) shows that the Qwen2.5-Coder-7B result is stable under decoding. Estimates on small subsets, such as the 55 Tier~1 records, are less precise, so we report 95\% Wilson intervals where we can. The 12 Qwen2.5-Coder-7B records without a verdict come from six functions, so these errors are not independent.

%% file: body/discussion.tex
\section{Discussion}
\label{sec:discussion}

Fine-tuning improved all three small open-weight models on the prompted task, though the gains differed substantially by model. Qwen2.5-Coder-7B's verdict accuracy rose from 4.8\% to 96.6\%; DeepSeek-Coder-6.7B rose from 1.1\% to 78.9\%, and StarCoder2-7B from 5.9\% to 43.7\%. The untuned models rarely returned a verdict in the required format. For Qwen2.5-Coder-7B, the lenient parser found a verdict in 150 of 355 base-model outputs, and 148 of those verdicts were correct. Fine-tuning therefore appears to help these models organize and express security evidence in the requested form. 

The grounding checks show what the fine-tuned models retained from the supplied evidence. Qwen2.5-Coder-7B cited 70 of 80 Tier~1 oracle sites, compared with 48 of 80 for the base model; all 70 cited lines were oracle sites. Its monologue header matched the record's track in all 343 outputs that included a header, and 165 of 172 VULNERABLE outputs with a verdict named the same CWE in the body and verdict line. These checks test whether the monologues reproduce selected details. In our review of 14 narrations, one Tier~2 explanation passed the checks despite failing to explain the vulnerability. The unchanged Tier~2 evidence-quotation rate after fine-tuning (57.4\% for both Qwen2.5-Coder-7B variants) also suggests that the gains are not uniform across evidence types.

The sampling results further distinguish the models. Qwen2.5-Coder-7B produced the same verdict in all five samples for every test record, with 96.6\% pass@1, pass@3, and pass@5. Its base model reached 16.7\% pass@1 and 58.9\% pass@5. DeepSeek-Coder-6.7B reached 87.2\% pass@1 and 96.1\% pass@5, while StarCoder2-7B reached 37.5\% and 89.3\%, respectively. For StarCoder2-7B, sampled pass@1 was below its 43.7\% greedy accuracy; for DeepSeek-Coder-6.7B, it was above its 78.9\% greedy accuracy. Thus, fine-tuning improved the sampled results for all three models, but the accuracy of a single output remained model- and decoding-dependent.

The five-function kernel case study offers an initial test beyond OGSM: secSemScan assigned the correct verdict and CWE to all five functions, while Semgrep produced no finding. The sample is too small to establish how often this will work on unseen vulnerabilities. The models' open weights and parameter counts make local deployment a relevant goal, especially where code cannot be sent to hosted services. However, we ran the experiments on NVIDIA L40S GPUs and did not measure memory use, latency, or power consumption on constrained hardware. Those measurements are needed to establish which devices can support the models in practice.

%% file: body/related_works.tex
\section{Related Work}
\label{sec:background}

\subsection{Static Analysis for Vulnerability Detection}

Static analyzers for C/C\texttt{++} detect vulnerabilities along two tracks. \emph{Taint analysis} follows untrusted data from a source (for example \texttt{copy\_from\_user}) to a dangerous sink (for example a \texttt{memcpy} size argument) and reports a vulnerability when no sanitizer lies on the path~\cite{yamaguchi2014modeling}. \emph{Lifetime analysis} follows heap objects from allocation through free to use and reports use-after-free (CWE-416) and double-free (CWE-415), which are temporal memory errors~\cite{szekeres2013sok}, and missing-free (CWE-401) when the sequence violates its invariants. CodeQL~\cite{codeql} implements both tracks but needs a compilable codebase, which excludes most isolated function snippets in vulnerability datasets. Semgrep~\cite{semgrep} matches patterns on source text with no build step, so it applies to such snippets. Rule-based analyzers find only what their rules encode, and six static C analyzers missed 47\% to 80\% of 192 known vulnerabilities in 27 real-world programs~\cite{lipp2022effectiveness}. secSemScan uses Semgrep as its \emph{oracle}, the analyzer whose traces supervise and check the model, and adds a learned model that aims to find vulnerabilities the rules miss.



\subsection{LLM-Based Vulnerability Detection}

Work on vulnerability detection has progressed from graph-based models and encoder transformers to decoder-only code models. Devign uses a graph neural network over code property graphs~\cite{zhou2019devign}, while encoder-based approaches such as LineVul apply transformers to vulnerability prediction~\cite{fu2022linevul}. More recent studies evaluate models such as CodeLlama, DeepSeek-Coder, and Qwen2.5-Coder, with fine-tuned models often outperforming earlier encoder-based approaches on standard benchmarks~\cite{ding2025primevul}. However, benchmark accuracy does not reliably predict performance on held-out CVEs or code from other domains~\cite{chakraborty2022reveal}; models can rely on dataset-specific cues such as function length, identifier names, and comments. Ahmed et al.~\cite{ahmed2026secvuleval} address a related evaluation gap with SecVulEval, a context-aware C/C\texttt{++} benchmark for statement-level vulnerability detection and localization, while secSemScan fine-tunes small models to narrate supplied analyzer traces or fix-diff evidence in a checkable format.

\subsection{Scope and Effectiveness of Open-weight LLMs}

Open-weight code LLMs run on local hardware, and training them on reasoning can raise their accuracy above that of much larger models. In R2Vul, Weyssow et al.~\cite{weyssow2025r2vul} have a teacher LLM write two reasoning examples per function, one for the true label and one for the flipped label, and train a student with reinforcement learning from AI feedback to prefer the former; across five programming languages, a 1.5B student is reported to exceed its 32B teacher and commercial LLMs in macro-F1. In VulLLM, Du et al.~\cite{du2024vulllm} add localization, built from the deleted lines of fix patches, and GPT-4-generated interpretation as auxiliary tasks to binary detection.

SemCoder~\cite{ding2024semcoder} trains a 6.7B code model to narrate a program's execution behavior in natural language (monologue reasoning), grounded in execution traces from executable Python, and the model outperforms GPT-3.5-turbo on HumanEval (79.3\% against 76.8\%). We take the term security monologue from this work and replace execution-trace grounding with static-oracle grounding, because taint propagation and object lifetime are static program facts, not runtime values.

Evidence on the limits of fine-tuning comes from CWE-Trace~\cite{cwetrace}, a diagnostic study and not a competing detector. It evaluates eight vanilla and fifteen LoRA fine-tuned LLMs on 834 Linux kernel samples spanning 74 CWEs under a temporal split that avoids training-data leakage, and the best configuration reaches 52.1\% detection accuracy and under 1.3\% CWE top-1 accuracy. The authors conclude that fine-tuning shifts a model's output threshold without producing the reasoning that reliable CWE classification requires.

R2Vul and VulLLM have a teacher LLM author the reasoning, and nothing requires it to name a taint or lifetime site that a tool can check independently of the model that produced it, although Du et al. verify the interpretations manually. secSemScan narrows the scope to two weakness families in C/C\texttt{++}, taint and lifetime, whose evidence is a small tuple of program sites, and requires each monologue to reproduce the oracle's named sites exactly, so reasoning can be judged claim by claim and not only by its final label.

\subsection{Neuro-Symbolic Approaches}

Pei et al.~\cite{pei2023can} evaluate LLM-predicted program invariants against reference invariants computed by Daikon from program executions. secSemScan compares the model's verdict with a fresh analyzer run at inference. However, in our evaluation, the prompt supplies the oracle trace (Section~\ref{sec:eval}), so this comparison does not test whether an analyzer independently confirms a claim produced by the model. IRIS~\cite{li2025iris}, GPTScan~\cite{sun2024gptscan}, LLift~\cite{li2024llift}, and LLMDFA~\cite{wang2024llmdfa} combine LLMs with program analysis for vulnerability detection. IRIS infers taint specifications for CodeQL on Java and reports its best results with hosted GPT-4. GPTScan uses a hosted GPT model to propose candidates in Solidity contracts, then confirms them with static analysis. LLift assesses static-analyzer warnings for use-before-initialization bugs in the Linux kernel, while LLMDFA has an LLM synthesize scripts that call a parser and an SMT solver. None of these four approaches trains a locally deployable model on evidence verified by a static analyzer. secSemScan instead fine-tunes a locally deployable model on monologues generated from Tier~1 analyzer traces and Tier~2 fix-diff evidence. We retain the LLM-generated narrations without post-generation filtering and evaluate whether their cited sites match the supplied evidence.

%% file: body/conclusion.tex
\section{Conclusion}
\label{sec:conclusion}


Fine-tuning helped three small open-weight code models produce more structured security monologues from supplied evidence. Qwen2.5-Coder-7B showed the largest gain, reaching 96.6\% verdict accuracy and citing 87.5\% of Tier~1 oracle sites; DeepSeek-Coder-6.7B reached 78.9\%, while StarCoder2-7B reached 43.7\%. These results suggest that fine-tuning can improve how small models express security evidence, but they do not establish independent vulnerability detection because the prompts included the verdict and evidence. The five-function kernel case study is encouraging but preliminary. Whether these models run effectively on compute-constrained hardware also remains open; evaluating memory, latency, and power use on target devices is an important next step.

%% file: main.bib
@inproceedings{ziegler2022productivity,
  title     = {Productivity Assessment of Neural Code Completion},
  author    = {Ziegler, Albert and Kalliamvakou, Eirini and Li, X. Alice and Rice, Andrew and Rifkin, Devon and Simister, Shawn and Sittampalam, Ganesh and Aftandilian, Edward},
  booktitle = {Proceedings of the 6th ACM SIGPLAN International Symposium on Machine Programming (MAPS)},
  year      = {2022}
}

@article{peng2023impact,
  title   = {The Impact of {AI} on Developer Productivity: Evidence from {GitHub Copilot}},
  author  = {Peng, Sida and Kalliamvakou, Eirini and Cihon, Peter and Demirer, Mert},
  journal = {arXiv preprint arXiv:2302.06590},
  year    = {2023}
}

@inproceedings{zhang2023prompt,
  title     = {Prompt-Enhanced Software Vulnerability Detection Using {ChatGPT}},
  author    = {Zhang, Chenyuan and Liu, Hao and Zeng, Jiutian and Yang, Kejing and Li, Yuhong and Li, Hui},
  booktitle = {Companion Proceedings of the International Conference on Software Engineering (ICSE Companion)},
  pages     = {276--277},
  year      = {2024}
}

@article{chakraborty2022reveal,
  title   = {Deep Learning Based Vulnerability Detection: Are We There Yet?},
  author  = {Chakraborty, Saikat and Krishna, Rahul and Ding, Yangruibo and Ray, Baishakhi},
  journal = {IEEE Transactions on Software Engineering},
  volume  = {48},
  number  = {9},
  pages   = {3280--3296},
  year    = {2022}
}

@inproceedings{he2023large,
  title     = {Large Language Models for Code: Security Hardening and Adversarial Testing},
  author    = {He, Jingxuan and Vechev, Martin},
  booktitle = {Proceedings of the ACM SIGSAC Conference on Computer and Communications Security (CCS)},
  pages     = {1865--1879},
  year      = {2023}
}

@inproceedings{ding2025primevul,
  title     = {Vulnerability Detection with Code Language Models: How Far Are We?},
  author    = {Ding, Yangruibo and Fu, Yanjun and Ibrahim, Omniyyah and Sitawarin, Chawin and Chen, Xinyun and Alomair, Basel and Wagner, David and Ray, Baishakhi and Chen, Yizheng},
  booktitle = {Proceedings of the IEEE/ACM International Conference on Software Engineering (ICSE)},
  year      = {2025}
}

@inproceedings{ullah2024llms,
  title     = {{LLMs} Cannot Reliably Identify and Reason About Security Vulnerabilities (Yet?): A Comprehensive Evaluation, Framework, and Benchmarks},
  author    = {Ullah, Saad and Han, Mingji and Pujar, Saurabh and Pearce, Hammond and Coskun, Ayse and Stringhini, Gianluca},
  booktitle = {Proceedings of the IEEE Symposium on Security and Privacy (S\&P)},
  pages     = {862--880},
  year      = {2024}
}

@inproceedings{liang2024large,
  title={A large-scale survey on the usability of ai programming assistants: Successes and challenges},
  author={Liang, Jenny T and Yang, Chenyang and Myers, Brad A},
  booktitle={Proceedings of the 46th IEEE/ACM international conference on software engineering},
  pages={1--13},
  year={2024}
}

@inproceedings{klemmer2024using,
  title={Using ai assistants in software development: A qualitative study on security practices and concerns},
  author={Klemmer, Jan H and Horstmann, Stefan Albert and Patnaik, Nikhil and Ludden, Cordelia and Burton Jr, Cordell and Powers, Carson and Massacci, Fabio and Rahman, Akond and Votipka, Daniel and Lipford, Heather Richter and others},
  booktitle={Proceedings of the 2024 on ACM SIGSAC Conference on Computer and Communications Security},
  pages={2726--2740},
  year={2024}
}

@inproceedings{rzanny2026decision,
  title={Decision-Makers’ Perceptions of Barriers to AI Coding Assistant Adoption—A Survey Study},
  author={Rzanny, Pawe{\l} and Ochodek, Miros{\l}aw and Staron, Miroslaw},
  booktitle={Euromicro Conference on Software Engineering and Advanced Applications},
  pages={162--178},
  year={2026},
  organization={Springer}
}

@inproceedings{khare2023understanding,
  title     = {Understanding the Effectiveness of Large Language Models in Detecting Security Vulnerabilities},
  author    = {Khare, Avishree and Dutta, Saikat and Li, Ziyang and Solko-Breslin, Alaia and Alur, Rajeev and Naik, Mayur},
  booktitle = {Proceedings of the IEEE Conference on Software Testing, Verification and Validation (ICST)},
  pages     = {103--114},
  year      = {2025}
}

@misc{weyssow2025r2vul,
  title        = {{R2Vul}: Learning to Reason about Software Vulnerabilities with Reinforcement Learning and Structured Reasoning Distillation},
  author       = {Weyssow, Martin and Yang, Chengran and Chen, Junkai and Widyasari, Ratnadira and Zhang, Ting and Huang, Huihui and Nguyen, Huu Hung and Tun, Yan Naing and Bui, Tan and Li, Yikun and Wei, Ang Han and Liauw, Frank and Ouh, Eng Lieh and Shar, Lwin Khin and Lo, David},
  howpublished = {arXiv preprint arXiv:2504.04699},
  year         = {2025}
}

@inproceedings{du2024vulllm,
  title     = {Generalization-Enhanced Code Vulnerability Detection via Multi-Task Instruction Fine-Tuning},
  author    = {Du, Xiaohu and Wen, Ming and Zhu, Jiahao and Xie, Zifan and Ji, Bin and Liu, Huijun and Shi, Xuanhua and Jin, Hai},
  booktitle = {Findings of the Association for Computational Linguistics: ACL 2024},
  pages     = {10507--10521},
  year      = {2024}
}

@inproceedings{sun2024gptscan,
  title={GPTScan: Detecting logic vulnerabilities in smart contracts by combining gpt with program analysis},
  author={Sun, Yuqiang and Wu, Daoyuan and Xue, Yue and Liu, Han and Wang, Haijun and Xu, Zhengzi and Xie, Xiaofei and Liu, Yang},
  booktitle={Proceedings of the IEEE/ACM 46th international conference on software engineering},
  pages={1--13},
  year={2024}
}

@inproceedings{li2025iris,
  title     = {{IRIS}: {LLM}-Assisted Static Analysis for Detecting Security Vulnerabilities},
  author    = {Li, Ziyang and Dutta, Saikat and Naik, Mayur},
  booktitle = {Proceedings of the International Conference on Learning Representations (ICLR)},
  year      = {2025}
}

@article{li2024llift,
  title   = {Enhancing Static Analysis for Practical Bug Detection: An {LLM}-Integrated Approach},
  author  = {Li, Haonan and Hao, Yu and Zhai, Yizhuo and Qian, Zhiyun},
  journal = {Proceedings of the ACM on Programming Languages},
  volume  = {8},
  number  = {OOPSLA1},
  pages   = {474--499},
  year    = {2024}
}

@inproceedings{wang2024llmdfa,
  title     = {{LLMDFA}: Analyzing Dataflow in Code with Large Language Models},
  author    = {Wang, Chengpeng and Zhang, Wuqi and Su, Zian and Xu, Xiangzhe and Xie, Xiaoheng and Zhang, Xiangyu},
  booktitle = {Advances in Neural Information Processing Systems (NeurIPS)},
  year      = {2024}
}

@inproceedings{ding2024semcoder,
  title     = {{SemCoder}: Training Code Language Models with Comprehensive Semantics Reasoning},
  author    = {Ding, Yangruibo and Peng, Jinjun and Min, Marcus J. and Kaiser, Gail and Yang, Junfeng and Ray, Baishakhi},
  booktitle = {Advances in Neural Information Processing Systems (NeurIPS)},
  year      = {2024}
}

@inproceedings{fu2022linevul,
  title     = {{LineVul}: A Transformer-based Line-Level Vulnerability Prediction},
  author    = {Fu, Michael and Tantithamthavorn, Chakkrit},
  booktitle = {Proceedings of the 19th International Conference on Mining Software Repositories (MSR)},
  pages     = {608--620},
  year      = {2022}
}

@inproceedings{hu2022lora,
  title     = {{LoRA}: Low-Rank Adaptation of Large Language Models},
  author    = {Hu, Edward J. and Shen, Yelong and Wallis, Phillip and Allen-Zhu, Zeyuan and Li, Yuanzhi and Wang, Shean and Wang, Lu and Chen, Weizhu},
  booktitle = {Proceedings of the International Conference on Learning Representations (ICLR)},
  year      = {2022}
}

@article{pearce2022asleep,
author = {Pearce, Hammond and Ahmad, Baleegh and Tan, Benjamin and Dolan-Gavitt, Brendan and Karri, Ramesh},
title = {Asleep at the Keyboard? Assessing the Security of GitHub Copilot’s Code Contributions},
year = {2025},
issue_date = {February 2025},
publisher = {Association for Computing Machinery},
address = {New York, NY, USA},
volume = {68},
number = {2},
issn = {0001-0782},
url = {https://doi.org/10.1145/3610721},
doi = {10.1145/3610721},
journal = {Commun. ACM},
month = jan,
pages = {96–105},
numpages = {10}
}

@inproceedings{diversevul,
  title     = {{DiverseVul}: A New Vulnerable Source Code Dataset for Deep Learning Based Vulnerability Detection},
  author    = {Chen, Yizheng and Ding, Zhoujie and Alowain, Lamya and Chen, Xinyun and Wagner, David},
  booktitle = {Proceedings of the 26th International Symposium on Research in Attacks, Intrusions and Defenses (RAID)},
  pages     = {654--668},
  year      = {2023}
}

@misc{pairvul,
  title={Vul-rag: Enhancing LLM-based vulnerability detection via knowledge-level rag},
  author={Du, Xueying and Zheng, Geng and Wang, Kaixin and Zou, Yi and Wang, Yujia and Deng, Wentai and Feng, Jiayi and Liu, Mingwei and Chen, Bihuan and Peng, Xin and others},
  journal={ACM Transactions on Software Engineering and Methodology},
  volume={35},
  number={10},
  pages={1--25},
  year={2026},
  publisher={ACM New York, NY}
}

@article{tsipenyuk2005seven,
  title={Seven pernicious kingdoms: A taxonomy of software security errors},
  author={Tsipenyuk, Katrina and Chess, Brian and McGraw, Gary},
  journal={IEEE Security \& Privacy},
  volume={3},
  number={6},
  pages={81--84},
  year={2005},
  publisher={IEEE}
}

@inproceedings{mvd,
  title     = {{MVD}: Memory-Related Vulnerability Detection Based on Flow-Sensitive Graph Neural Networks},
  author    = {Cao, Sicong and Sun, Xiaobing and Bo, Lili and Wu, Rongxin and Li, Bin and Tao, Chuanqi},
  booktitle = {Proceedings of the 44th IEEE/ACM International Conference on Software Engineering (ICSE)},
  pages     = {1456--1468},
  year      = {2022}
}

@inproceedings{szekeres2013sok,
  title     = {{SoK}: Eternal War in Memory},
  author    = {Szekeres, Laszlo and Payer, Mathias and Wei, Tao and Song, Dawn},
  booktitle = {Proceedings of the IEEE Symposium on Security and Privacy (S\&P)},
  year      = {2013}
}

@inproceedings{yamaguchi2014modeling,
  title     = {Modeling and Discovering Vulnerabilities with Code Property Graphs},
  author    = {Yamaguchi, Fabian and Golde, Nico and Arp, Daniel and Rieck, Konrad},
  booktitle = {Proceedings of the IEEE Symposium on Security and Privacy (S\&P)},
  year      = {2014}
}

@misc{hui2024qwen25coder,
  title        = {Qwen2.5-Coder Technical Report},
  author       = {Hui, Binyuan and Yang, Jian and Cui, Zeyu and Yang, Jiaxi and Liu, Dayiheng and Zhang, Lei and Liu, Tianyu and Zhang, Jiajun and Yu, Bowen and Dang, Kai and others},
  howpublished = {arXiv preprint arXiv:2409.12186},
  year         = {2024}
}

@misc{semgrep,
  title        = {Semgrep: Fast, Customizable, and Language-Agnostic Static Analysis},
  author       = {{Semgrep, Inc.}},
  howpublished = {\url{https://semgrep.dev}},
  year         = {2024}
}

@inproceedings{zhou2019devign,
  title     = {Devign: Effective Vulnerability Identification by Learning Comprehensive Program Semantics via Graph Neural Networks},
  author    = {Zhou, Yaqin and Liu, Shangqing and Siow, Jingkai and Du, Xiaoning and Liu, Yang},
  booktitle = {Advances in Neural Information Processing Systems (NeurIPS)},
  volume    = {32},
  year      = {2019}
}

@inproceedings{pei2023can,
  title     = {Can Large Language Models Reason About Program Invariants?},
  author    = {Pei, Kexin and Bieber, David and Shi, Kensen and Sutton, Charles and Yin, Pengcheng},
  booktitle = {Proceedings of the 40th International Conference on Machine Learning (ICML)},
  volume    = {202},
  pages     = {27496--27520},
  year      = {2023}
}

@misc{codeql,
  title        = {{CodeQL}: Semantic Code Analysis Engine},
  author       = {{GitHub, Inc.}},
  howpublished = {\url{https://codeql.github.com}},
  year         = {2024}
}

@misc{cwetrace,
  title        = {Calibration Without Comprehension: Diagnosing the Limits of Fine-Tuning {LLM}s for Vulnerability Detection in Systems Software},
  author       = {Zibaeirad, Arastoo and Vieira, Marco},
  howpublished = {arXiv preprint arXiv:2606.20502},
  year         = {2026},
 
}

@article{robbes2026agentic,
  title={Agentic much? adoption of coding agents on github},
  author={Robbes, Romain and Matricon, Th{\'e}o and Degueule, Thomas and Hora, Andre and Zacchiroli, Stefano},
  journal={ACM Transactions on Software Engineering and Methodology},
  year={2026},
  publisher={ACM New York, NY}
}

@article{zhang2026ccrab,
  title={Code review agent benchmark},
  author={Zhang, Yuntong and Pan, Zhiyuan and Yusuf, Imam Nur Bani and Ruan, Haifeng and Shariffdeen, Ridwan and Roychoudhury, Abhik},
  journal={arXiv preprint arXiv:2603.23448},
  year={2026}
}

@inproceedings{lipp2022effectiveness,
  title={An empirical study on the effectiveness of static C code analyzers for vulnerability detection},
  author={Lipp, Stephan and Banescu, Sebastian and Pretschner, Alexander},
  booktitle={Proceedings of the 31st ACM SIGSOFT international symposium on software testing and analysis},
  pages={544--555},
  year={2022}
}

@article{guo2024deepseekcoder,
  title={DeepSeek-Coder: when the large language model meets programming--the rise of code intelligence},
  author={Guo, Daya and Zhu, Qihao and Yang, Dejian and Xie, Zhenda and Dong, Kai and Zhang, Wentao and Chen, Guanting and Bi, Xiao and Wu, Yifan and Li, YK and others},
  journal={arXiv preprint arXiv:2401.14196},
  year={2024}
}

@article{lozhkov2024starcoder2,
  title={Starcoder 2 and the stack v2: The next generation},
  author={Lozhkov, Anton and Li, Raymond and Allal, Loubna Ben and Cassano, Federico and Lamy-Poirier, Joel and Tazi, Nouamane and Tang, Ao and Pykhtar, Dmytro and Liu, Jiawei and Wei, Yuxiang and others},
  journal={arXiv preprint arXiv:2402.19173},
  year={2024}
}

@inproceedings{ahmed2026secvuleval,
  title={SecVulEval: Context-Aware Benchmarking of LLMs for Vulnerability Detection},
  author={Ahmed, Md Basim Uddin and Harzevili, Nima Shiri and Shin, Jiho and Pham, Hung Viet and Wang, Song},
  booktitle={Proceedings of the 3rd ACM International Conference on AI-Powered Software},
  pages={388--396},
  year={2026}
}
